\documentclass[prb,twocolumn,showpacs, superscriptaddress,floatfix]{revtex4}

\usepackage{bm}
\usepackage{graphicx,color}
\usepackage{dcolumn} 

\newcommand{\rbk}[1]{\left( #1 \right)}

\newcommand{\sbk}[1]{\left[ #1 \right]}

\newcommand{\retn}{\nonumber \\ }

\newcommand{\arrcase}[1]{\left\{ \begin{array}{ll} #1 \end{array}\right.}
\newcommand{\vc}[1]{\bm{#1}}
\newcommand{\MARU}[1]{{\ooalign{\hfil#1\/\hfil\crcr\raise .167ex\hbox{\mathhexbox20D}}}}

\begin{document}

\title{Chemical Reaction between Single Hydrogen Atom and Graphene}

\author{Atsushi ITO}
 \email{ito.atsushi@nifs.ac.jp}
 \affiliation{Department of Physics, Graduate School of Science, Nagoya University, Chikusa, Nagoya 464-8602, Japan.}
\author{Hiroaki NAKAMURA}%
 \email{nakamura.hiroaki@nifs.ac.jp}
\affiliation{National Institute for Fusion Science, Oroshi-cho 322-6, Toki 509-5292, Japan.}
\author{Arimichi TAKAYAMA}%
 \email{takayama@nifs.ac.jp}
\affiliation{National Institute for Fusion Science, Oroshi-cho 322-6, Toki 509-5292, Japan.}

\date{\today} 

\begin{abstract}
We study chemical reaction between a single hydrogen atom and a graphene, which is the elemental reaction between hydrogen and graphitic carbon materials.
In the present work, classical molecular dynamics simulation is used with modified Brenner's empirical bond order potential.
The three reactions, that is, absorption reaction, reflection reaction and penetration reaction, are observed in our simulation.
Reaction rates depend on the incident energy of the hydrogen atom and the graphene temperature.
The dependence can be explained by the following mechanisms:
(1) The hydrogen atom receives repulsive force by $\pi$--electrons in addition to nuclear repulsion.
(2) Absorbing the hydrogen atom, the graphene transforms its structure to the ``overhang'' configuration such as $\textit{sp}^3$ state.
(3) The hexagonal hole of the graphene is expanded during the penetration of the hydrogen atom. 
\end{abstract}

\maketitle

\section{Introduction}

Chemical reaction between hydrogen and graphite is a basic process of the chemical sputtering between plasma and a divertor plate in plasma confinement experiments. \cite{Nakano,Roth,Roth2,Mech,LHD}
The divertor plate composed of graphite tiles or carbon fiber composites is bombarded with hydrogen plasma.
Eroding the carbon wall, hydrogen ions  yield $\mathrm{H}_2$ and several sorts of hydrocarbon molecules, i.e., $\mathrm{ CH_x, C_2 H_x}$ and so on.
The hydrocarbon molecules misbehave as impurities for plasma confinement experiments.
In order to reduce the hydrocarbon impurities in the plasma, we need to understand the erosion mechanism of  carbon walls and the creation mechanism of hydrocarbon molecules.
However, these mechanisms are not clarified yet.
We, therefore, reveal the chemical reaction between hydrogen and graphite by computer simulation.
 
In the present work, we study the chemical reaction between a single hydrogen atom and a graphene with classical molecular dynamics (CMD) simulation.
It is reasonable to consider that ions of the hydrogen plasma combine with electrons and become neutral atoms before interacting with the carbon wall.
Therefore, we select a neutral hydrogen atom as an injected particle.
A graphene is the elemental component of graphitic carbon materials. \cite{Boehm}
The chemical reaction between the single hydrogen atom and the graphene, therefore,  is regarded as the elemental reaction between hydrogen and various graphitic carbon materials.

We measured only an absorption rate in our previous work, \cite{Ito} where multi-layer graphite was treated.
In the present work, we evaluate a reflection rate and a penetration rate in addition to the absorption rate.
We also obtain the incident hydrogen energy dependence and the graphene temperature dependence of each reaction rate.
In \S \ref{ss:SimMethod}, the simulation model and method are described.
Simulation results are shown in \S \ref{ss:Result}.
We discuss the feature of the chemical reaction between the hydrogen atom and the graphene in \S \ref{ss:Discuss}.
This paper is concluded with summary in \S \ref{ss:Summary}.
In addition, we note the modification of Brenner's reactive empirical bond order potential in Appendix.

\section{Simulation Method}\label{ss:SimMethod}

In the present work, we adapt CMD simulation with the \textit{NVE} condition, in which the number of particle, volume and total energy are conserved.
The second order symplectic integration \cite{Suzuki} is used to solve the time evolution of the equation of motion. 
The time step is $5 \times 10^{-18} \mathrm{~s}$.
We use a modified Brenner's reactive empirical bond order (REBO) potential: \cite{Brenner}
\begin{eqnarray}
	U \equiv \sum_{i,j>i} \sbk{V_{[ij]}^\mathrm{R}( r_{ij} )
		 - \bar{b}_{ij}(\{r\},\{\theta^\mathrm{B}\},\{\theta^\mathrm{DH}\}) V_{[ij]}^\mathrm{A}(r_{ij}) }, \nonumber \\
	\label{eq:model_rebo}
\end{eqnarray}
where $r_{ij}$ is the distance between the $i$-th and the $j$-th atoms. 
The functions $V_{[ij]}^{\mathrm{R}}$ and $V_{[ij]}^{\mathrm{A}}$ represent repulsion and attraction, respectively.
The function $\bar{b}_{ij}$ generates multi--body force.
We show details of the modified Brenner's REBO potential in Appendix.

Figure \ref{fig:simbox} shows the present simulation model.
The hydrogen atom is injected into the graphene composed of 160 carbon atoms.
The center of mass of the graphene is set to the origin of coordinates.
The surface of the graphene is parallel to the $x$--$y$ plane.
The size of the graphene is 2.13 nm $\times$ 1.97 nm.
The graphene has no lattice defects and no crystal edges due to periodic boundary condition toward $x$ and $y$ directions.
The graphene temperature is defined by
\begin{eqnarray}
	T \equiv \frac{2}{3 N k_{\mathrm{b}}} \sum_i^{\mathrm{ carbon}} \frac{\vc{p}_i^2}{2 m_i},
\end{eqnarray}
where $\vc{p}_i$ and $m_i$ are the momentum and the mass of the $i$--th carbon atom, respectively.
$N$ is the number of carbon atoms and $k_{\mathrm{ b}}$ is the Boltzmann constant.
The symbol $\sum_i^{\mathrm{carbon}}$ denotes the summation over the carbon atoms.
The carbon atoms obey the Maxwell--Boltzmann distribution in the initial state of the simulation.

The hydrogen atom is injected parallel to the $z$ axis from $z = 4$~\AA.
We repeat 200 simulations where the $x$ and $y$ coordinates of injection points are set at random.
As a result, we obtain the histograms, which give reaction rates.
The incident energy $E_{\mathrm{ I}}$ determines the initial momentum $\vc{p}_\mathrm{H}(0)=(0,0,p_0)$ of the hydrogen atom as 
\begin{eqnarray}
	p_0 = \sqrt{2 m_{\mathrm{ H}} E_{\mathrm{ I}}}, \label{eq:inip}
\end{eqnarray}
where $m_{\mathrm{ H}}$ is the mass of the hydrogen atom.

\begin{figure}
	\centering
	\resizebox{85mm}{!}{\includegraphics{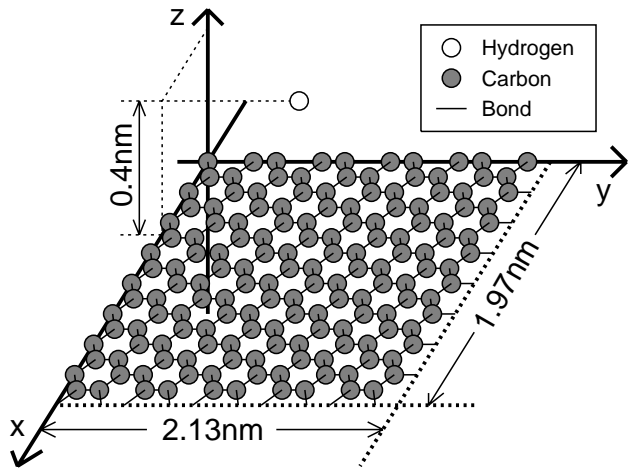}} \\
	\caption{Simulation model. There are 160 carbon atoms and an injected hydrogen atom.}
	\label{fig:simbox}
\end{figure}

\section{Results}\label{ss:Result}

Three kinds of reactions between the single hydrogen atom and the graphene are observed in our CMD simulation.
They are absorption reaction, reflection reaction and penetration reaction.
The properties of the reactions are described in the following.

\subsection{Dynamics of three reactions}

\begin{figure}
	\centering
	\resizebox{\linewidth}{!}{\includegraphics{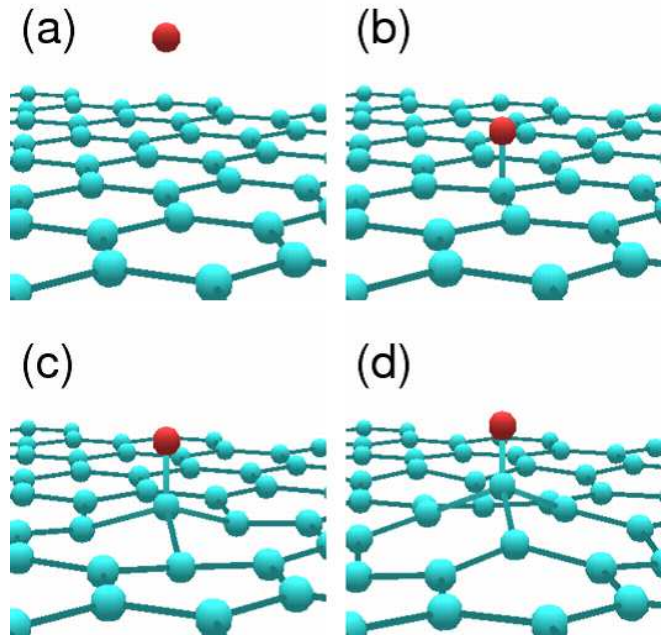}} \\
	\caption{Snapshots of the absorption reaction.
		(a) The hydrogen atom with the incident energy of 3 eV is injected.
		(b) The hydrogen atom makes a covalent bond with the nearest carbon atom.
		(c) The nearest carbon atom is pulled out of the graphene and the ``overhang'' configuration appears.
		(d) The atoms are relaxed to the stable structure with oscillation.}
	\label{fig:XX1}
\end{figure}

In the absorption reaction, the hydrogen atom and the nearest carbon atom are bound by a new covalent bond.
The nearest carbon atom is pulled out of the surface of the graphene as a $\textit{sp}^3$ configuration (Fig. \ref{fig:XX1}).
We call this phenomenon ``overhang''.
The hydrogen atom remains  above the nearest carbon atom while oscillating.
In the reflection reaction, the graphene reflects the incident hydrogen atom to the region of $z > 0$.
In the penetration reaction, the incident hydrogen atom passes through the graphene and goes away to the region of $z < 0$.
It is observed that the graphene expands the hexagonal hole while the hydrogen atom is penetrating.

\subsection{Incident energy dependence of reaction rates}
\begin{figure}
	\centering
		\resizebox{\linewidth}{!}{\includegraphics{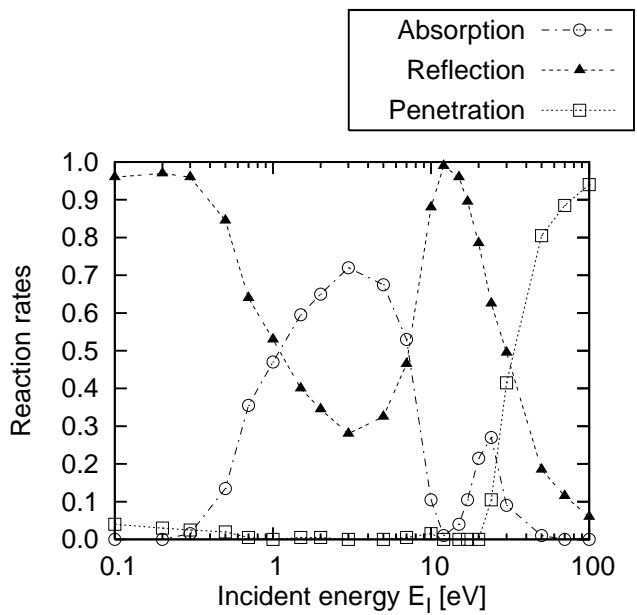}} \\
	\caption{Incident energy dependence of the absorption, the reflection and the penetration rates.
	  Dash-dotted line with	open circle, long-dashed line with filled triangle, and short-dashed line with	square denote the absorption, the reflection and the penetration rates, respectively.}
	\label{fig:XX2}
\end{figure}

Figure \ref{fig:XX2} shows the incident energy dependence of  each reaction rate in the case that the initial graphene temperature is 300 K.
Three kinds of reactions dominate in different incident energy $E_{\mathrm{ I}}$ respectively.
In the case of $E_{\mathrm{ I}} < 1 \mathrm{~eV}$, almost all of the incident hydrogen atoms are  reflected.
For $1 \mathrm{~eV} < E_{\mathrm{ I}} < 7 \mathrm{~eV}$, the absorption reaction becomes dominant.
The reflection reaction becomes dominant again for $7 \mathrm{~eV} < E_{\mathrm{I}} < 30  \mathrm{~eV}$.
The penetration reaction behaves as the dominant process for $E_{\mathrm{I}} > 30 \mathrm{~eV}$.
It is also observed that the absorption rate has the small peak around 24 eV in Fig. \ref{fig:XX2}.

\subsection{Graphene temperature dependence of reaction rates}

\begin{figure*}
\begin{tabular}{cccc}
		\resizebox{0.28\linewidth}{!}{\includegraphics{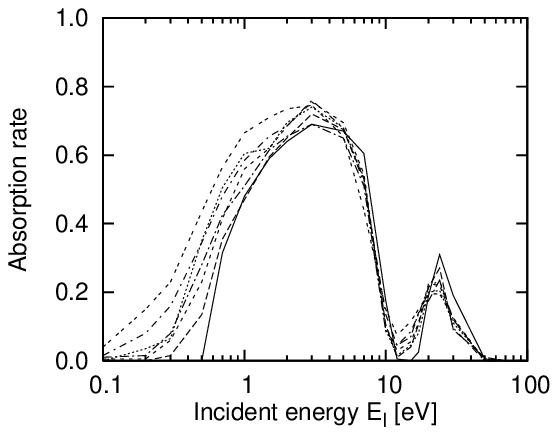}} &
		\resizebox{0.28\linewidth}{!}{\includegraphics{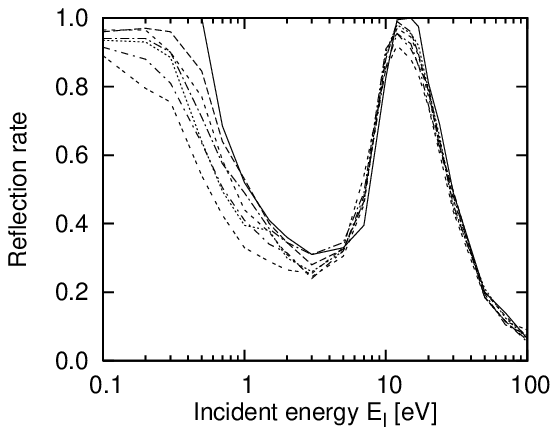}} &
		\resizebox{0.28\linewidth}{!}{\includegraphics{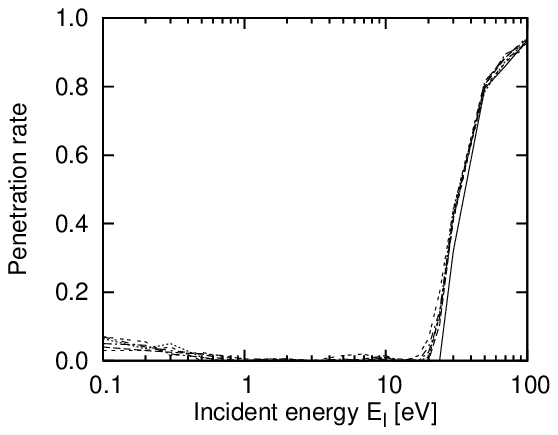}} &
		\resizebox{0.1\linewidth}{!}{\includegraphics{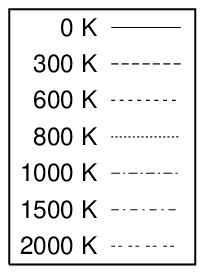}} \\
	\end{tabular}
\caption{Graphene temperature dependence of the absorption, the reflection and the penetration rates.
We varied the graphene temperature such as 0 K, 300 K, 600 K, 800 K, 1000 K, 1500 K and 2000 K.}
\label{fig:XX3}
\end{figure*}

We also investigate the initial graphene temperature dependence of reaction rates (Fig. \ref{fig:XX3}).
As the initial graphene temperature rises, the absorption rate tends to broaden to the region of low incident energy.
In contrast, the reflection rate drops.
However, the graphene temperature hardly affects the penetration rate.

\section{Discussions}\label{ss:Discuss}

The three reactions were observed in the present simulation.
The incident energy dependence and the graphene temperature dependence of the three reaction rates are also obtained.
The similar incident energy dependence of three reactions was observed in the system of multi--layer graphite and a large amount of hydrogen atoms. \cite{Nakamura}
It is, therefore, important to understand the mechanism of the chemical reaction between a single hydrogen atom and a single graphene to argue atomic--scale processes in various systems composed of hydrogen and graphitic carbon materials.

\subsection{Two kinds of repulsive force}

\begin{figure}
	\centering
		\resizebox{85mm}{!}{\includegraphics{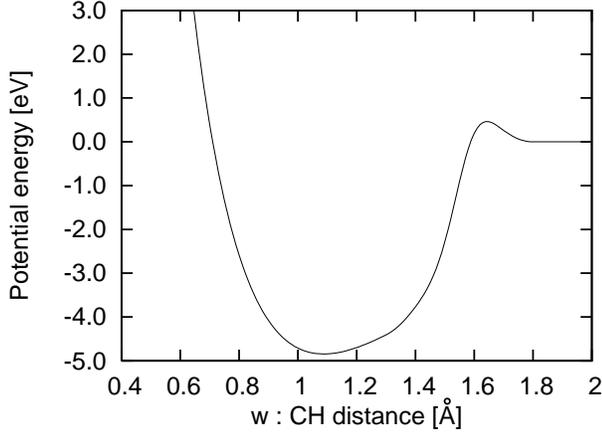}}
	\caption{The potential energy between an incident hydrogen atom and a graphene.
	The potential energy is calculated by the modified REBO potential model.
	The position of the incident hydrogen atom is set to be $(x_0, y_0, z_0 +w)$, where $(x_0, y_0, z_0 )$ is the position of the nearest carbon atom and $w > 0$.
	The other carbons are relaxed to the total potential energy minimum state for each $w$. }
	\label{fig:XX4}
\end{figure}

It was observed that the reflection reaction dominates in two ranges of the incident energy, i.e., $E_{\mathrm{I}} < 1 \mathrm{~eV}$~and $7 \mathrm{~eV} < E_{\mathrm{I}} < 30 \mathrm{~eV}$~in Fig. \ref{fig:XX2}.
From this fact,  it is deduced that two kinds of repulsive force work between the incident hydrogen atom and the graphene.
To prove it, we plot the potential energy between the hydrogen atom and the graphene in Fig. \ref{fig:XX4}, where the hydrogen atom is located just above the nearest carbon atom at the distance $w$ and the other carbon atoms are relaxed.
From Fig. \ref{fig:XX4}, we can confirm the existence of two kinds of repulsive force between the incident hydrogen atom and the graphene.
The first repulsive force for  $w < 1.0$~\AA~ is due to the repulsive term $V_{[ij]}^{\mathrm{R}}$ in Eq.  (\ref{eq:model_rebo}) and corresponds to nuclear repulsion.
The second repulsive force for $1.6 \textrm{~\AA} < w < 1.8 \textrm{~\AA}$~is derived from the multi--body force in the term $\bar{b}_{ij}$ in Eq. (\ref{eq:model_rebo}).
The existence of the second repulsive force was also confirmed by \textit{ab--initio} calculations. \cite{Sha,Jeloaicia}
It was considered that $\pi$--electrons over the graphene generate the second repulsive force.
The energy height of the potential wall of the second repulsive force is estimated to be  about 0.5~eV.
The hydrogen atom with the incident energy of 0.5~eV or more, therefore, can enter the region that $ w < 1.6 \textrm{~\AA}$, in which the other mechanism derives the absorption reaction and the penetration reaction.
The mechanisms for $E_{\mathrm{ I}} > 0.5 \mathrm{~eV}$~and $ w < 1.6 \textrm{~\AA}$~are described in the subsequent subsections.
Thus, the reflection rate starts to decrease at $E_{\mathrm{ I}} \sim 0.5 \mathrm{~eV}$~in Fig. \ref{fig:XX2}.

\subsection{Absorption mechanism}

\begin{figure}
\centering
		\resizebox{85mm}{!}{\includegraphics{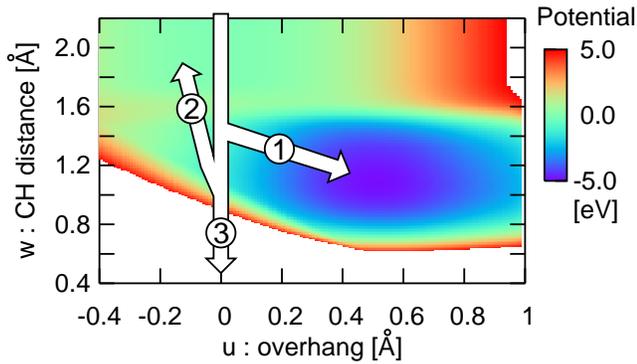}} \\
\caption{Potential energy contour in the $u$ and $w$ parameter space.
The parameters $u$ and $w$ are indicated in Fig. \ref{fig:XX6}.
The potential energy is higher than $5$~eV in the white area.
The trident arrow represents the trajectory of each reaction.
The numbers 1, 2 and 3 correspond to the absorption,
 the reflection, and the penetration reactions, respectively.
}	\label{fig:XX5}
\end{figure}

\begin{figure}
\centering
		\resizebox{85mm}{!}{\includegraphics{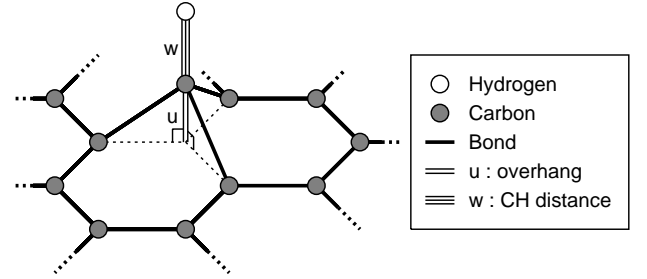}}
\caption{Schematic picture of the ``overhang'' configuration.
White and gray circles represent the incident hydrogen atom and the carbon atoms, respectively.
Single lines represent covalent bonds.
The parameter $w$ is the distance between the incident hydrogen atom  and the nearest carbon atom. 
The length $u$ of the double line represents the height of the overhang from the surface of the graphene.
}
\label{fig:XX6}
\end{figure}

Figure \ref{fig:XX5} shows the potential energy contour in the $(u, w)$ parameter space.
Here $u$ is the height of the nearest carbon atom from the surface of the graphene  and $w$ is the distance between the hydrogen atom and the nearest carbon atom as shown in Fig. \ref{fig:XX6}.
There is the minimum potential point at $(u,w)$ = (0.5~\AA, 1.1~\AA), which corresponds to the ``overhang'' configuration. 
This analysis by the potential energy contour claims that the ``overhang'' configuration is the most stable state.

The trajectory of the absorption reaction is represented by  arrow \MARU{1} in Fig. \ref{fig:XX5}.
The initial state corresponds to the point of $(u, w)$ = (0~\AA, 4~\AA).
The hydrogen atom and the graphene start interaction around $(u, w) \sim (0\textrm{~\AA}, 1.8\textrm{~\AA})$.
The arrow \MARU{1} shows that, with tumbling down the slope of the potential energy contour, the state falls into the potential minimum point $(u,w)\sim($0.5 \AA, 1.1 \AA$)$, which indicates the ``overhang'' configuration.

Until the hydrogen atom overcomes the second repulsive force, the graphene cannot transform its structure to the ``overhang'' configuration.
Thereby, the incident energy of 0.5 eV, which is corresponding to the energy height of the potential wall of the second repulsive force, is the lower limit to occur the absorption reaction.
The absorption rate rises from $E_{\mathrm{ I}} \sim 0.5$~ eV in contrast to the reflection rate by the second repulsive force and has a peak at $E_{\mathrm{ I}} = 3$~ eV.

\subsection{Reflection mechanism}

In the reflection reaction for $7 \mathrm{~eV} < E_{\mathrm{ I}} < 30 \mathrm{~eV}$, the incident hydrogen atom bounds back from the potential wall $V_{[ij]}^{\mathrm{ R}}$ in Eq. ({\ref{eq:model_rebo}}), which is represented by the region of $w < 1.0 \textrm{~\AA}$~in Fig. \ref{fig:XX4} and the white region in Fig. \ref{fig:XX5}.
After bounding, the hydrogen atom goes away from the graphene without connecting the nearest carbon atom.
Therefore, the graphene keeps the flat sheet configuration and does not transform its structure to the ``overhang'' configuration.
The trajectory of the reflection reaction by $V_{[ij]}^{\mathrm{ R}}$ is drawn  as arrow \MARU{2} in Fig. \ref{fig:XX5}.

Here, to make clear a distinction between the absorption reaction and the reflection reaction for $E_{\mathrm{ I}} > 1 \mathrm{~eV}$, we introduce two typical time lengths $\Delta t_u$ and $\Delta t_w$.
The time length $\Delta t_u$ is defined as the time length necessary for the graphene absorbing the hydrogen atom to transform its structure from the ``flat sheet'' configuration to the ``overhang'' configuration.
Strictly speaking, $\Delta t_u$ depends on a lot of parameters, for example, the incident energy and the incident position of the hydrogen atom, the graphene temperature, and so on.
But, to estimate the typical time length $\Delta t_u$, we  consider  the only simple overhang process, which is represented by the following trajectory in the parameter space Fig. \ref{fig:XX5}: the configuration of the atoms is transformed from the start point that $(u,w)=(0 \textrm{~\AA},~1.1 \textrm{~\AA})$ to the end point that $(u,w)=(0.5 \textrm{~\AA},~1.1 \textrm{~\AA})$~along a straight line $w=1.1 \textrm{~\AA}$.
We, moreover,  assume that the initial velocities of all the atoms are zero.
The potential function $U_{\mathrm{oh}}$ along the above path is approximated by the following harmonic oscillator:
\begin{equation}
		U_{\mathrm{oh}} (u) = \frac{(m_{\mathrm{ C}} + m_{\mathrm{ H}}) \omega^2 (u-u_0)^2 }{2} - U_0,
		\label{eq:deltau1}
\end{equation}
where we use, as the mass, the sum of $m_{\mathrm{ C}} = 12.0$~amu and $m_{\mathrm{ H}}= 1.00 $~amu, because the hydrogen atom and the nearest carbon atom move as a rigid body in our assumption that $w$ is fixed to the constant length of 1.1~\AA. 
From Fig. \ref{fig:XX5}, we have the potential minimum point $u_0 = 0.5$~\AA, the minimum potential-energy $U_0 = 4.84$ eV, and $\omega= 1.45 \times 10^{14} \mathrm{~s}^{-1}$.
Thus, we can estimate $\Delta t_u$ as follows:
\begin{equation}
		\Delta t_u = \frac{1}{4} \rbk{\frac{2\pi}{\omega}} = 1.08 \times 10^{-14} {\mathrm{~s}}.
		\label{eq:deltau3}
\end{equation}
This estimated value of $\Delta t_u$ is comparable to the CMD simulation result $\Delta t_u^{\mathrm{ sim.}} \sim 1.28 \times 10^{-14} \mathrm{~s}$, which was obtained under the condition that the degrees of the freedom except the parameter $u$ are fixed to the initial values.

The other time length $\Delta t_w$ is defined as the time length in which the hydrogen atom can stay in the region that  $w < 1.6$~\AA.
To estimate $\Delta t_w$, we adapt the alternative	assumption that  the hydrogen atom moves as a free particle for $w > 0.9$~\AA~ along the straight line of $u = 0$~\AA~ and collides with the potential wall $V_{[ij]}^{\mathrm{ R}}$ at $(u,w) = ( 0$~\AA, 0.5~\AA).
From this assumption and Eq. (\ref{eq:inip}), $\Delta t_w$ is given by 
\begin{eqnarray}
		\Delta t_w = \frac{2 l m_{\mathrm{ H}}}{p_0} = l \sqrt{\frac{2 m_{\mathrm{ H}}}{E_{\mathrm{ I}}}},
		\label{eq:deltaw1}
\end{eqnarray}
where $l = (1.6 - 0.9)\textrm{~\AA} = 0.7 \textrm{~\AA}$. 
From Eq. (\ref{eq:deltaw1}), it is obtained that $\Delta t_w$ is proportional to $1/\sqrt{ E_{\mathrm{ I}} } $. 
On the other hand, Eq. (\ref{eq:deltau3}) shows that $\Delta t_u$  is independent of $E_{\mathrm{ I}}$.

Comparing these time length, we consider the following two cases.
In the first case that $\Delta t_w > \Delta t_u$, the hydrogen atom connects with the nearest carbon and the graphene transforms its structure to the ``overhang'' configuration, before the hydrogen atom escapes to the region that $w > 1.6$~\AA.
The hydrogen atom, therefore, is absorbed by the graphene.
As the incident energy increases, $\Delta t_w$ becomes smaller than $\Delta t_u$.
In this case $\rbk{\Delta t_u > \Delta t_w = l \sqrt{{2 m_{\mathrm{ H}}}/{E_{\mathrm{ I}}}} }$, the hydrogen atom escapes  before the graphene traps the hydrogen atom.
This process is regarded as  the reflection reaction.
We can derive  the following condition necessary for the reflection reaction:
\begin{eqnarray}
	   E_{\mathrm{ I}} > \frac{2 l^2 m_{\mathrm{ H}}}{{\Delta t_u}^2} = 0.84 {\mathrm{~eV}}.
	   \label{eq:deltaw2}
\end{eqnarray}
The incident energy which satisfies the condition {$\Delta t_w = \Delta t_u$} is estimated as {$E_{\mathrm{ I}}^{\mathrm{ sim}} \sim 1$}~eV in our CMD simulation where the degrees of freedom except {$w$} are fixed to the initial values.
By comparison between $E_{\mathrm{ I}}^{\mathrm{ sim}}$ and the condition Eq. {(\ref{eq:deltaw2})}, it is considered that the our assumption is proper.
In the above discussion, the hydrogen atom is located on the vertical axis over the nearest carbon atom.
However, In the present simulation, the hydrogen atom seldom exists just above the nearest carbon atom, because the $x$ and $y$ coordinates of the incident hydrogen atom are set at random.
Thereby, the repulsive force by $V_{[ij]}^{\mathrm{ R}}$ becomes weaker than that of the potential energy contour in Fig. \ref{fig:XX5}.
The time length $\Delta t_w$ becomes, therefore, longer than the estimated value of Eq. (\ref{eq:deltaw1}).
The hydrogen atom which deviates from the vertical axis over the nearest carbon atom needs higher incident energy than the estimated value of Eq. (\ref{eq:deltaw2}).
Consequently, the incident energy of 0.84 eV in Eq. (\ref{eq:deltaw2}) is the lower limit to occur the reflection reaction by $V_{[ij]}^{\mathrm{ R}}$.

\subsection{Penetration mechanism}

\begin{figure}
	\centering
		\resizebox{85mm}{!}{\includegraphics{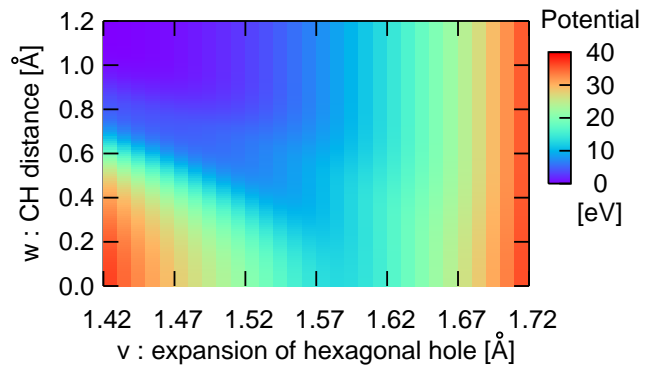}} \\
	\caption{Potential energy contour in the $v$ and $w$ parameter space.
	The parameters $v$ and $w$ are indicated in Fig. \ref{fig:XX8}.
	The Interaction force does not act on the hydrogen atom and the graphene when the state is located on $w > 1.11$~\AA.}
	\label{fig:XX7}
\end{figure}

\begin{figure}
\centering
		\resizebox{85mm}{!}{\includegraphics{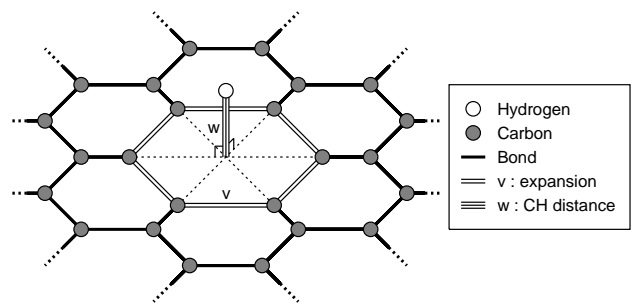}}
\caption{Expansion of the hexagonal hole of the graphene.
White circle represents the incident hydrogen atom.
Gray circles represent carbon atoms.
Single lines represent covalent bonds.
The parameter $w$ is the distance between the incident hydrogen atom  and the surface of the graphene.
The double lines represent expanding covalent bonds,  and the parameter $v$ is the length of a side of the hexagonal hole.
}
\label{fig:XX8}
\end{figure}

We describe the dynamics of the penetration reaction.
We notice for the present simulation that the graphene expands the hexagonal hole during the penetration of the hydrogen atom.
Figure \ref{fig:XX7} shows the potential energy contour with two parameters, i.e., the distance $w$ and the length $v$ of the side of the hexagonal hole (See Fig. \ref{fig:XX8}).
We note that the hydrogen atom is located above the center of the hexagonal hole unlike the layout of Fig. \ref{fig:XX6}.
The C--C bond length of the stable graphene structure is 1.42~\AA.
The interaction force acts on the hydrogen atom and the graphene in $w < 1.11$~\AA.
There is the potential minimum region of 0 eV in the area that $v = 1.42$~\AA~ and $w > 1.11$~\AA, which is the incident state of the hydrogen atom.
If the size of the hexagonal hole $v$ is fixed to 1.42 \AA, the energy height of potential wall is 38 eV at $(v, w) = (1.42\textrm{~\AA},0\textrm{~\AA})$.
In this case, the hydrogen atom needs the incident energy of 38 eV or more to penetrate the graphene.
However, the penetration reaction with the incident energy of less than 38 eV is observed in the present simulation result Fig. \ref{fig:XX2}.
The difference between the estimation and the simulation result is explained by the expansion mechanism of the hexagonal hole of the graphene.
If carbon atoms move along the bottom of the potential energy valley in Fig. \ref{fig:XX7}, the parameter $v$ increases from 1.42~\AA~to 1.58~\AA~with decreasing $w$.
Thus, the hexagonal hole is expanded as the hydrogen atom approaches the graphene.
As a consequence, the energy height of the potential wall is lowered to 13 eV at $(v, w) = (1.58\textrm{~\AA},0\textrm{~\AA})$.
This expansion lets the hydrogen atom penetrate in the incident energy of less than 38 eV.

Here, we indicate that the carbon atoms can expand the hexagonal hole before reflecting the hydrogen atom with the incident energy of 13 eV.
We define $\Delta t'_w$ as the time length for the hydrogen atom to approach the graphene from $w = 1.11$~\AA.
The time length $\Delta t'_w$ is given by
\begin{eqnarray}
		\Delta t'_w = \frac{l' m_{\mathrm{ H}}}{p_0} = l' \sqrt{\frac{m_{\mathrm{ H}}}{2E_{\mathrm{ I}}}} = 2.18 \times 10^{-15} {\mathrm{~s}}.
		\label{eq:pen1}
\end{eqnarray}
where $l' = 1.11$~\AA~ and $E_I$ is set to 13 eV.
On the other hand, the potential energy around $v = 1.58$~\AA, where $w$ is fixed to 0~\AA, is approximated by the following harmonic oscillator:
\begin{eqnarray}
		U_{\mathrm{ hole}} (v) = \frac{m_{\mathrm{ hole}} \omega'^2 (v-v_0)^2 }{2} + U'_0,
		\label{eq:pen2}
\end{eqnarray}
where $m_{\mathrm{ hole}} = 6 m_{\mathrm{ C}}$, $v_0 = 1.58$~\AA, $U'_0 = 13$ eV and $\omega' = 5.21 \times 10^{14} \mathrm{~s}^{-1}$ from the potential energy contour Fig. \ref{fig:XX7}.
For this approximation, we obtain the time length $\Delta t_v$ to accomplish the expansion of the hexagonal hole as follows:
\begin{eqnarray}
		\Delta t_v = \frac{1}{4} \rbk{\frac{2\pi}{\omega'}} = 3.01 \times 10^{-15} \mathrm{~s}.
		\label{eq:pen3}
\end{eqnarray}
Both $\Delta t'_w$ and $\Delta t_v$ are on the same order of femtosecond.
In addition, the time length $\Delta t'_w$ of Eq. (\ref{eq:pen1}) becomes practically larger than $2.18 \times 10^{-15} \mathrm{~s}$ because of deceleration due to repulsion.
Therefore, the carbon atoms can expand the hexagonal hole in response to the approach of the hydrogen atom.

Next, we consider the small peak of the absorption rate at $E_{\mathrm{I}} = 24 \mathrm{~eV}$, at which the hydrogen atom has enough incident energy to penetrate the graphene.
The incident energy of the hydrogen atom diffuses into the graphene.
Therefore, the hydrogen atom has no longer the necessary incident energy to escape from the graphene.
From the energy diffusion, it is understood that the peak of the absorption reaction at 24 eV is caused by the hydrogen atom absorption on the reverse side of the graphene.
The absorption reaction on the reverse side was confirmed in the present simulation.
As long as the hydrogen atom is absorbed, the graphene transforms its structure to the ``overhang'' configuration where the nearest carbon atom is pulled into the reverse side of the graphene.

\subsection{Graphene temperature dependence of reaction rates}

The graphene temperature dependence of reaction rates is significant at low incident energies in Fig. \ref{fig:XX3}.
As the graphene temperature is raised, the absorption rate increases and the reflection rate decreases for $E_{\mathrm{I}} < 1$~eV.
The maximum temperature 2000 K, which  corresponds to $0.26$~eV as kinetic energy per a carbon atom, is comparable to the energy height of the potential wall of the second repulsive force of 0.5 eV.
If the nearest carbon atom has kinetic energy, the relative momentum between the hydrogen atom and the nearest carbon atom becomes larger than the initial momentum of the hydrogen atom $p_0$ in Eq. (\ref{eq:inip}).
In the case of high graphene temperature, therefore, we substitute the relative momentum for $p_0$ and can perform similar estimation to the preceding subsections.
As a result, the absorption rate increases and the reflection rate decreases as the graphene temperature is raised.
By comparison  energy order, the penetration rate is insensitive to the graphene temperature.

\section{Summary}\label{ss:Summary}

By the CMD simulation with modified Brenner's REBO potential model, we demonstrated the chemical reaction between the single hydrogen atom and the single graphene, which can be regarded as the elemental reaction between hydrogen and graphitic carbon materials.
We observed the three processes, which are the absorption, the reflection and the penetration reactions.
The dominant reaction is replaced according to the incident energy for  $0.1 \mathrm{~eV} \leq E_{\mathrm{ I}} \leq 100 \mathrm{~eV}$.
We discussed the characteristic interactions between the hydrogen atom and the graphene with potential energy.
The hydrogen atom receives the repulsive force not only by nuclei of carbon atoms but also by $\pi$--electrons over the surface of the graphene.
These two kinds of repulsive force cause the two reflection mechanisms.
When the hydrogen atom is absorbed, the graphene is transformed from flat sheet configuration to ``overhang'' configuration.
By comparison between the typical time length of the overhang transformation and the time length during the hydrogen atom's stay, we can clarify the difference between the absorption reaction and the reflection reaction for $E_{\mathrm{ I}} > 0.5$~eV.
In the penetration reaction, the incident hydrogen atom goes through the hexagonal hole of the graphene and the graphene expands the hexagonal hole, simultaneously.
The expansion lowers the energy height of the potential wall by nuclei of the carbon atoms, which accounts for starting the penetration reaction at $E_{\mathrm{ I}} = 13$~eV in Fig. \ref{fig:XX2}.
In addition, we investigated the graphene temperature dependence of reaction rates.
As the graphene temperature rises, the absorption rate increases and the reflection rate decreases for low incident energy.
The cause of the graphene temperature dependence is that the kinetic energy of the nearest carbon atom is comparable to the energy height of the potential wall by $\pi$--electrons.

\section*{ACKNOWLEDGMENTS}

The authors acknowledge Professor Shinji Tsuneyuki and Dr. Yoshihide Yoshimoto for helpful comments, and Dr. Noriyasu Ohno for stimulating discussions.
Numerical simulations are carried out by use of the Plasma Simulator at National Institute for Fusion Science.
The work is supported partly by Grand-in Aid for Exploratory Research (C), 2006, No.~17540384 from the Ministry of Education, Culture, Sports, Science and Technology and  partly by the National Institutes of Natural Sciences undertaking for Forming Bases for Interdisciplinary and International Research through Cooperation Across Fields of Study, and Collaborative Research Programs (No. NIFS06KDAT012 and No. NIFS06KTAT029).

\appendix*

\section{MODIFIED BRENNER'S REBO POTENTIAL MODEL}
We note here the review of Brenner's  reactive empirical bond order (REBO) potential \cite{Brenner} and our modification points.
This potential model follows in the wake of Morse potential, \cite{Morse}  Abell potential \cite{Abell} and Tersoff potential. \cite{Tersoff1,Tersoff2}

The potential function $U$ is defined by
\begin{eqnarray}
	U \equiv \sum_{i,j>i} \sbk{V_{[ij]}^\mathrm{R}( r_{ij} )
		 - \bar{b}_{ij}(\{r\},\{\theta^\mathrm{B}\},\{\theta^\mathrm{DH}\}) V_{[ij]}^\mathrm{A}(r_{ij}) }, \nonumber \\
	\label{eq:p7}
\end{eqnarray}
where $r_{ij}$ is the distance between the $i$--th and the $j$--th atoms.
The bond angle $\theta_{jik}^\mathrm{B}$ is the angle between the line segment which starts at the $i$--th atom and ends at the $j$--th atom  and the line segment which starts at the $i$--th atom and ends at the $k$--th atom,  as follows:
\begin{eqnarray}
	\cos\theta_{jik}^\mathrm{B} = \frac{\rbk{\bm{x}_j - \bm{x}_i} \cdot \rbk{\bm{x}_k - \bm{x}_i}}
		{r_{ij} r_{ik}},
	\label{eq:p4}
\end{eqnarray}
where $\bm{x}_i$ is the position coordinate of the $i$--th atom and $r_{ij}$ is the distance between the $i$-th and the $j$-th atoms..
The dihedral angle $\theta_{kijl}^\mathrm{DH}$ is the angle between the triangle formed by the $j$--th, the $i$--th and the $k$--th atoms and the triangle formed by the $i$--th, the $j$--th and the $l$--th atoms.
The cosine function of $\theta_{kijl}^\mathrm{DH}$ is given by 
\begin{eqnarray}
	\cos\theta_{kijl}^\mathrm{DH} = \frac{(\bm{x}_i - \bm{x}_k) \times (\bm{x}_j - \bm{x}_i)}{r_{ik}r_{ji}}
		\cdot \frac{(\bm{x}_j-\bm{x}_i) \times (\bm{x}_l-\bm{x}_j)}{r_{ji}r_{lj}}. \nonumber\\
\end{eqnarray}

The repulsive function $V_{[ij]}^\mathrm{R}( r_{ij})$ and the attractive function $V_{[ij]}^\mathrm{A}( r_{ij})$ are defined by
\begin{eqnarray}
	V_{[ij]}^\mathrm{R}( r_{ij}) &\equiv& f_{[ij]}^\mathrm{c}(r_{ij}) \rbk{1+\frac{Q_{[ij]}}{r_{ij}}}
		A_{[ij]} \exp\rbk{-\alpha_{[ij]} r_{ij}}, \nonumber\\ \\
	V_{[ij]}^\mathrm{A}( r_{ij}) &\equiv& f_{[ij]}^\mathrm{c}(r_{ij}) \sum_{n=1}^3 B_{n[ij]} \exp\rbk{-\beta_{n[ij]} r_{ij}}.
\end{eqnarray}
The square bracket such as $[ij]$ means that each function or each parameter depends only on the species of the $i$--th and the $j$--th atoms,  for example $V_\mathrm{CC}^\mathrm{R}$, $V_\mathrm{HH}^\mathrm{R}$ and $V_\mathrm{CH}^\mathrm{R}$~($= V_\mathrm{HC}^\mathrm{R}$).
The coefficients $Q_{[ij]}$, $A_{[ij]}$, $\alpha_{[ij]}$, $B_{n[ij]}$~ and $\beta_{n[ij]}$~  are given by Table \ref{tb:tb2body}.

\begin{table*}
\caption{The parameters for the repulsive function $V_{[ij]}^\mathrm{R}$ and the attractive function $V_{[ij]}^\mathrm{A}$.
 They depend on the species of the $i$--th and the $j$--th atoms.}
\label{tb:tb2body}
\begin{ruledtabular}
\begin{tabular}{cccc}
	~ & \multicolumn{3}{c}{$[ij]$} \\
	\cline{2-4}
	Parameter & CC & HH & CH or HC \\
	\hline
	$Q_{[ij]}$ & 0.3134602960833 \AA & 0.370471487045 \AA & 0.340775728 \AA \\
	$A_{[ij]}$ & 10953.544162170 eV & 32.817355747 eV & 149.94098723 eV \\
	$\alpha_{[ij]}$ & 4.7465390606595 $\textrm{\AA}^{-1}$ & 3.536298648 $\textrm{\AA}^{-1}$ & 4.10254983 $\textrm{\AA}^{-1}$ \\
	$B_{1[ij]}$ & 12388.79197798 eV & 29.632593 eV & 32.3551866587 eV \\
	$B_{2[ij]}$ & 17.56740646509 eV & 0 eV & 0 eV \\
	$B_{3[ij]}$ & 30.71493208065 eV & 0 eV & 0 eV \\
	$\beta_{1[ij]}$ & 4.7204523127 $\textrm{\AA}^{-1}$ & 1.71589217 $\textrm{\AA}^{-1}$ & 1.43445805925 $\textrm{\AA}^{-1}$ \\
	$\beta_{2[ij]}$ & 1.4332132499 $\textrm{\AA}^{-1}$ & 0 $\textrm{\AA}^{-1}$ & 0 $\textrm{\AA}^{-1}$ \\
	$\beta_{3[ij]}$ & 1.3826912506 $\textrm{\AA}^{-1}$ & 0 $\textrm{\AA}^{-1}$ & 0 $\textrm{\AA}^{-1}$ \\
\end{tabular}
\end{ruledtabular}
\end{table*}

The cutoff function $f_{[ij]}^\mathrm{c}(r_{ij})$ determines effective ranges of the covalent bond between the $i$--th and the $j$--th atoms.
Two atoms are bound with the covalent bond if the distance $r_{ij}$ is shorter than $D_{[ij]}^\mathrm{min}$.
Two atoms are not bound with the covalent bond if the distance $r_{ij}$ is longer than $D_{[ij]}^\mathrm{max}$.
The cutoff function $f_{[ij]}^\mathrm{c}(r_{ij})$ connects the above two states smoothly as
\begin{eqnarray}
	f_{[ij]}^\mathrm{c}(x) \equiv \left\{ \begin{array}{l} 
				1 \hspace{10em} (x \leq D_{[ij]}^\mathrm{min}), \\
				\frac{1}{2}\sbk{1+\cos(\pi \frac{x - D_{[ij]}^\mathrm{min}}{D_{[ij]}^\mathrm{max} - D_{[ij]}^\mathrm{min}})} \\
				\hspace{7em} (D_{[ij]}^\mathrm{min} < x \leq D_{[ij]}^\mathrm{max}), \\
				0 \hspace{10em} (x > D_{[ij]}^\mathrm{max}). \\ 
		\end{array}\right. \retn
\end{eqnarray}
The constants $D_{[ij]}^\mathrm{min}$ and $D_{[ij]}^\mathrm{max}$ depend on the species of the two atoms (Table \ref{tb:tbDminmax}).
The cutoff function $f_{[ij]}^\mathrm{c}(r_{ij})$ distinguishes the presence of the covalent bond  between the $i$--th and the $j$--th atoms.

\begin{table}
\caption{The constants for the cutoff function $f_{[ij]}^\mathrm{c}(r_{ij})$.
 They depend on the species of the $i$--th and the $j$--th atoms.
}
\label{tb:tbDminmax}
\begin{ruledtabular}
\begin{tabular}{@{\hspace{2.0em}}ccc@{\hspace{2.0em}}}
	[ij] & $D_{[ij]}^\mathrm{min}$~(\AA) & $D_{[ij]}^\mathrm{max}$~(\AA) \\
	\hline
	CC & 1.7 & 2.0 \\
	CH & 1.3 & 1.8 \\
	HH & 1.1 & 1.7 \\
\end{tabular}
\end{ruledtabular}
\end{table}

The potentials $V_{[ij]}^\mathrm{R}$ and $V_{[ij]}^\mathrm{A}$ in Eq. (\ref{eq:p7}) generate two--body force, because both are the function of the only distance $r_{ij}$.
The multi--body force is used instead of the effect of an electron orbital.
In this model, $\bar{b}_{ij}(\{r\},\{\theta^\mathrm{B}\},\{\theta^\mathrm{DH}\})$ in Eq. (\ref{eq:p7})   gives multi--body force and is defined by 
\begin{eqnarray}
	&&\bar{b}_{ij}(\{r\},\{\theta^\mathrm{B}\},\{\theta^\mathrm{DH}\}) \retn
	 &&\indent \equiv \frac{1}{2} \Big[ b_{ij}^{\sigma-\pi}(\{r\},\{\theta^\mathrm{B}\}) + b_{ji}^{\sigma-\pi}(\{r\},\{\theta^\mathrm{B}\}) \Big] \retn
		&& \indent \indent+ \Pi_{ij}^\mathrm{RC}(\{r\}) + b_{ij}^\mathrm{DH}(\{r\},\{\theta^\mathrm{DH}\}).
	\label{eq:p2}
\end{eqnarray}
The first term $\frac{1}{2}\sbk{\cdots}$ generates three--body force  except the effect of $\pi$--electrons.
The second term $\Pi_{ij}^\mathrm{RC}$ in Eq. (\ref{eq:p2}) represents the influence of radical energetics and $\pi$--bond conjugation. \cite{Brenner}
The third term $b_{ij}^\mathrm{DH}(\{r\},\{\theta^\mathrm{DH}\})$ in Eq.  (\ref{eq:p2})   derives four--body force in terms of dihedral angles.
These functions are composed of the production of cutoff functions $f_{[ij]}^\mathrm{c}(r_{ij})$.
Five-- or more--body force are generated during chemical reaction.

\begin{table}
\caption{The parameters for the sixth order spline function $G_\mathrm{C}(\cos\theta_{jik}^\mathrm{B})$.}
\label{tb:tb6spGC}
\begin{ruledtabular}
\begin{tabular}{ccccc}
	$\cos\theta_{jik}^\mathrm{B}$ & $G_\mathrm{C}$ & $G_\mathrm{C}'$ & $G_\mathrm{C}''$ & $G_\mathrm{C}^{(3)}$ \\
	\hline
	$-1$ & $-$0.001 & 0.10400 & 0 & 0 \\
	$-1/2$ & 0.05280 & 0.170 & 0.370 & $-$5.232 \\ 
	$\cos(109.47^{\circ})$ & 0.09733 & 0.400 & 1.980 & 41.6140 \\ 
	1 & 8.0 & 0.23622 & $-$166.1360 & --- \\ 
\end{tabular}
\end{ruledtabular}
\end{table}

\begin{table}[t]
\caption{The parameters for the sixth order spline function $\gamma_\mathrm{C}(\cos\theta_{jik}^\mathrm{B})$.}
\label{tb:tb6spGammaC}
\begin{ruledtabular}
\begin{tabular}{ccccc}
	$\cos\theta_{jik}^\mathrm{B}$ & $\gamma_\mathrm{C}$ & $\gamma_\mathrm{C}'$ & $\gamma_\mathrm{C}''$ & $\gamma_\mathrm{C}^{(3)}$ \\
	\hline
	$\cos(109.47^{\circ})$ & 0.09733 & 0.400 & 1.980 & $-$9.9563027\\
	1 & 1.0 & 0.78 & $-$11.3022275  & --- \\ 
\end{tabular}
\end{ruledtabular}
\end{table}

\begin{table}
\caption{The parameters for the sixth order spline function $G_\mathrm{H}(\cos\theta_{jik}^\mathrm{B})$.
	The parameters are determined under $\cos\theta_{jik}^\mathrm{B} = 0$.}
\label{tb:tb6spGH}
\begin{ruledtabular}
\begin{tabular}{@{\hspace{5.0em}}lr@{\hspace{5.0em}}}
	Parameter & Value\hspace{1.0em} \\
	\hline
	$G_\mathrm{H}(0)$       &19.06510\\
	$G_\mathrm{H}'(0)$      & 1.08822\\
	$G_\mathrm{H}''(0)$     &-1.98677\\
	$G_\mathrm{H}^{(3)}(0)$ & 8.52604\\
	$G_\mathrm{H}^{(4)}(0)$ &-6.13815\\
	$G_\mathrm{H}^{(5)}(0)$ &-5.23587\\
	$G_\mathrm{H}^{(6)}(0)$ & 4.67318\\
\end{tabular}
\end{ruledtabular}
\end{table}

The function $b_{ij}^{\sigma-\pi}(\{r\},\{\theta^\mathrm{B}\})$ in Eq.  (\ref{eq:p2}) is defined by
\begin{eqnarray}
	b_{ij}^{\sigma-\pi}(\{r\},\{\theta^\mathrm{B}\})
	&\equiv& \Big[ 1 + \sum_{k\neq i,j} f_{[ij]}^\mathrm{c} (r_{ij})
		 \tilde{G}_i(\cos\theta_{jik}^\mathrm{B}) e^{\lambda_{[ijk]}} \retn
	 &&\indent + P_{[ij]}(N_{ij}^\mathrm{H},N_{ij}^\mathrm{C})\Big]^{-\frac{1}{2}}. \label{eq:p1}
\end{eqnarray}
The function $\tilde{G}_i$ in Eq. (\ref{eq:p1}) depends on the species of the $i$--th atom.
If $\cos\theta_{jik}^\mathrm{B} > \cos(109.47^{\circ})$ and the $i$--th atom is carbon, $\tilde{G}_i$ is defined by
\begin{eqnarray}
	\tilde{G}_i(\cos\theta_{jik}^\mathrm{B}) &\equiv& \sbk{1-Q_\mathrm{c}(M_i^\mathrm{t})}G_\mathrm{C}(\cos\theta_{jik}^\mathrm{B}) \retn
			 &&+ Q_\mathrm{c}(M_i^\mathrm{t})\gamma_\mathrm{C}(\cos\theta_{jik}^\mathrm{B}).
		\label{eq:p8}
\end{eqnarray}
If $\cos\theta_{jik}^\mathrm{B} \leq \cos(109.47^{\circ})$ and the $i$--th atom is carbon, $\tilde{G}_i$ is defined by
\begin{eqnarray}
	\tilde{G}_i(\cos\theta_{jik}^\mathrm{B}) \equiv G_\mathrm{C}(\cos\theta_{jik}^\mathrm{B}).
\end{eqnarray}
And, if the $i$--th atom is hydrogen, $\tilde{G}_i$ is defined by
\begin{eqnarray}
		\tilde{G}_i(\cos\theta_{jik}^\mathrm{B}) \equiv G_\mathrm{H}(\cos\theta_{jik}^\mathrm{B}).
\end{eqnarray}
Here $G_\mathrm{C}$, $\gamma_\mathrm{C}$ and $G_\mathrm{H}$ are the sixth order polynomial spline functions.
Though the spline function $\tilde{G}_i$ needs seven coefficients, the only six coefficients are written in Brenner's paper. \cite{Brenner}
We determine the seven coefficients in table \ref{tb:tb6spGC}, \ref{tb:tb6spGammaC} and \ref{tb:tb6spGH}, respectively.
The function $Q_\mathrm{c}$ and the coordination number $M_i^\mathrm{t}$ in Eq. (\ref{eq:p8}) are defined by
\begin{eqnarray}
	Q_\mathrm{c}(x) \equiv \arrcase{1 & \rbk{x \leq 3.2}, \\
			\frac{1}{2}\sbk{1 + \cos\rbk{2\pi\rbk{x-3.2}} } & \rbk{3.2 < x \leq 3.7}, \\
			0 & \rbk{x > 3.7}, \\} \retn
\end{eqnarray}
\begin{eqnarray}
	M_i^\mathrm{t} \equiv \sum_{k \neq i} f_{[ik]}^\mathrm{c}(r_{ik}).
\end{eqnarray}

The constant $\lambda_{[ijk]}$ in Eq. (\ref{eq:p1}) is  a weight to modulate a strength of three--body force, which  depends on the species of the $i$--th, the $j$--th and the $k$--th atoms.
In comparison with Brenner's former potential, \cite{Bernner90} we set constants $\lambda_{[ijk]}$ as follows:
\begin{eqnarray}
	\lambda_\mathrm{HHH} &=& 4.0, \\
	\lambda_\mathrm{CCC} &=& \lambda_\mathrm{CCH} = \lambda_\mathrm{CHC} = \lambda_\mathrm{HCC} \nonumber\\
	 &=& \lambda_\mathrm{HHC} = \lambda_\mathrm{HCH} = \lambda_\mathrm{CHH} = 0.
\end{eqnarray}

\begin{table}
\caption{Parameters for the bicubic spline function $P_{[ij]}(N_{ij}^\mathrm{H},N_{ij}^\mathrm{C})$.
 The parameters which are not denoted are zero.
}
\label{tb:tbPij}
\begin{ruledtabular}
\begin{tabular}{@{\hspace{2.0em}}cd@{\hspace{2.0em}}}
	$P_{[ij]}(N_{ij}^\mathrm{H},N_{ij}^\mathrm{C})$ & \multicolumn{1}{c}{\hspace{3.0em}Value} \\ 
	\hline
$P_\mathrm{CC}(1, 1)$& 0.003026697473481 \\ 
$P_\mathrm{CC}(2, 0)$& 0.007860700254745 \\ 
$P_\mathrm{CC}(3, 0)$& 0.016125364564267 \\ 
$P_\mathrm{CC}(1, 2)$& 0.003179530830731 \\ 
$P_\mathrm{CC}(2, 1)$& 0.006326248241119 \\ 
$P_\mathrm{CH}(1, 0)$&  0.2093367328250380 \\ 
$P_\mathrm{CH}(2, 0)$& -0.064449615432525 \\ 
$P_\mathrm{CH}(3, 0)$& -0.303927546346162 \\ 
$P_\mathrm{CH}(0, 1)$&  0.01 \\ 
$P_\mathrm{CH}(0, 2)$& -0.1220421462782555 \\ 
$P_\mathrm{CH}(1, 1)$& -0.1251234006287090 \\ 
$P_\mathrm{CH}(2, 1)$& -0.298905245783 \\ 
$P_\mathrm{CH}(0, 3)$& -0.307584705066 \\ 
$P_\mathrm{CH}(1, 2)$& -0.3005291724067579 \\ 
\end{tabular}
\end{ruledtabular}
\end{table}

The function $P_{[ij]}$ in Eq. (\ref{eq:p1}) is required  in the case that molecules forms solid structure.
The function $P_{[ij]}$ is the bicubic spline function whose coefficients depend on the species of the $i$--th and the $j$--th atoms (Table \ref{tb:tbPij}).
The parameters $N_{ij}^\mathrm{H}$ and $N_{ij}^\mathrm{C}$ are, respectively, the number of hydrogen atoms and the number of carbon atoms bound by the $i$--th atom as follows:
\begin{eqnarray}
N_{ij}^\mathrm{H} \equiv \sum_{k \neq i,j}^\mathrm{hydrogen} f_{[ik]}^\mathrm{c}(r_{ik}), \\
N_{ij}^\mathrm{C} \equiv \sum_{k \neq i,j}^\mathrm{carbon} f_{[ik]}^\mathrm{c}(r_{ik}).
\end{eqnarray}

\begin{table}
\caption{Parameters for the tricubic spline function $F_{[ij]}$.
 The parameters which are not denoted are zero.
 The function $F_{[ij]}$ satisfies the following rules:
 $F_{[ij]}(N_1,N_2,N_3)=F_{[ij]}(N_2,N_1,N_3)$,
 $\partial_{N_1} F_{[ij]}(N_1,N_2,N_3)  = \partial_{N_1} F_{[ij]}(N_2,N_1,N_3)$,
 $F_{[ij]}(N_1,N_2,N_3)=F_{[ij]}(3,N_2,N_3)$~ if $N_1>3$,
 and $F_{[ij]}(N_1,N_2,N_3)=F_{[ij]}(N_1,N_2,5)$~ if $N_3>5$,
 where $\partial_{N_i} \equiv \partial / \partial N_i$.
}
\label{tb:tbFij}
\begin{ruledtabular}
\begin{tabular}{ccccl}
	~ & \multicolumn{3}{c}{Variables} & ~ \\ 
	\cline{2-4}
	Function & $N_1$ & $N_2$ & $N_3$ & \hspace{2.0em}Value \\ 
	\hline
	$F_\mathrm{CC}(N_1,N_2,N_3)$& 1& 1& 1& \hspace{0.4525em} 0.105000 \\ 
	~ & 1& 1& 2& $-$0.0041775 \\ 
	~ & 1& 1& 3 to 5& $-$0.0160856 \\ 
	~ & 2& 2& 1& \hspace{0.4525em} 0.09444957 \\ 
	~ & 2& 2& 2& \hspace{0.4525em} 0.04632351 \\ 
	~ & 2& 2& 3& \hspace{0.4525em} 0.03088234 \\ 
	~ & 2& 2& 4& \hspace{0.4525em} 0.01544117 \\ 
	~ & 2& 2& 5& \hspace{0.4525em} 0.0 \\ 
	~ & 0& 1& 1& \hspace{0.4525em} 0.04338699 \\ 
	~ & 0& 1& 2& \hspace{0.4525em} 0.0099172158 \\ 
	~ & 0& 2& 1& \hspace{0.4525em} 0.0493976637 \\ 
	~ & 0& 2& 2& $-$0.011942669 \\ 
	~ & 0& 3& 1 to 5& $-$0.119798935 \\ 
	~ & 1& 2& 1& \hspace{0.4525em} 0.0096495698 \\ 
	~ & 1& 2& 2& \hspace{0.4525em} 0.030 \\ 
	~ & 1& 2& 3& $-$0.0200 \\ 
	~ & 1& 2& 4 to 5& $-$0.030133632 \\ 
	~ & 1& 3& 2 to 5& $-$0.124836752 \\ 
	~ & 2& 3& 1 to 5& $-$0.044709383 \\ 
	$\partial_{N_1} F_\mathrm{CC}(N_1,N_2,N_3)$ & 2& 1& 1& $-$0.052500 \\ 
	~ & 2& 1& 3 to 5& $-$0.054376 \\ 
	~ & 2& 3& 1& \hspace{0.4525em} 0.0 \\ 
	~ & 2& 3& 2 to 5& \hspace{0.4525em} 0.062418 \\ 
	$\partial_{N_3} F_\mathrm{CC}(N_1,N_2,N_3)$ & 2& 2& 4& $-$0.006618 \\ 
	~ & 1& 1& 2& $-$0.060543 \\ 
	~ & 1& 2& 3& $-$0.020044 \\ 
	$F_\mathrm{HH}(N_1,N_2,N_3)$ & 1& 1& 1& \hspace{0.4525em} 0.249831916 \\ 
	$F_\mathrm{CH}(N_1,N_2,N_3)$ & 0& 2& 3 to 5& $-$0.0090477875161288110 \\ 
	~ & 1& 3& 1 to 5& $-$0.213 \\ 
	~ & 1& 2& 1 to 5& $-$0.25 \\ 
	~ & 1& 1& 1 to 5& $-$0.5 \\ 
\end{tabular}
\end{ruledtabular}
\end{table}

The second term $\Pi_{ij}^\mathrm{RC}$ in Eq.  (\ref{eq:p2}) is defined by a tricubic spline function $F_{[ij]}$ as
\begin{eqnarray}
	\Pi_{ij}^\mathrm{RC}(\{r\}) \equiv F_{[ij]} (N_{ij}^\mathrm{t},N_{ji}^\mathrm{t},N_{ij}^\mathrm{conj}),
	\label{eq:p6}
\end{eqnarray}
where the variables are defined by
\begin{eqnarray}
	N_{ij}^\mathrm{t} &\equiv& \sum_{k \neq i,j} f_{[ik]}^\mathrm{c}(r_{ik}), \\
	N_{ij}^\mathrm{conj} &\equiv& 1 + \sum_{k(\neq i,j)}^\mathrm{carbon} f_{[ik]}^\mathrm{c}(r_{ik})C_\mathrm{N}(N_{ki}^\mathrm{t})\retn
		 &&\indent + \sum_{l(\neq j,i)}^\mathrm{carbon} f_{[jl]}^\mathrm{c}(r_{jl})C_\mathrm{N}(N_{lj}^\mathrm{t}), 
		\label{eq:p9}
\end{eqnarray}
with
\begin{eqnarray}
	C_\mathrm{N}(x) \equiv \arrcase{1 & (x \leq 2), \\
				\frac{1}{2}\sbk{1+\cos(\pi (x - 2))} & (2 < x \leq 3), \\
				0 & (x > 3). \\ } \retn
\end{eqnarray}
The second and the third terms of the right hand of Eq. (\ref{eq:p9}) are not squared.
We note that they are squared in Brenner's original formulation. \cite{Brenner}
By this modification, a numerical error becomes smaller than Brenner's formation.
Table \ref{tb:tbFij} shows the revised coefficients for $F_{[ij]}$.

The third term $b_{ij}^\mathrm{DH}(\{r\},\{\theta^\mathrm{DH}\})$ in Eq. (\ref{eq:p2}) is defined by
\begin{eqnarray}
	&&b_{ij}^\mathrm{DH}(\{r\},\{\theta^\mathrm{DH}\}) \equiv T_{[ij]}(N_{ij}^\mathrm{t},N_{ji}^\mathrm{t},N_{ij}^\mathrm{conj}) \retn
		&& \indent \times \sbk{\sum_{k\neq i,j} \sum_{l\neq j,i}
			\rbk{1-\cos^2\theta_{kijl}^\mathrm{DH}}f_{[ik]}^\mathrm{c}(r_{ik})f_{[jl]}^\mathrm{c}(r_{jl})}, \retn
\end{eqnarray}
where $T_{[ij]}$ is a tricubic spline function  and has the same variables as $F_{[ij]}$ in Eq. (\ref{eq:p6}).
The coefficients for $T_{[ij]}$ is also revised due to the modified $N_{ij}^\mathrm{conj}$~(Table \ref{tb:tbTij}).
In the present simulation, the function $T_{[ij]}$ becomes $T_\mathrm{CC}(2,2,5)$ for a perfect crystal graphene, and becomes $T_\mathrm{CC}(2,2,3)$ or $T_\mathrm{CC}(2,2,4)$ when a hydrogen atom is absorbed.

\begin{table}[t]
\caption{Parameters for the tricubic spline function $T_\mathrm{CC}$.
 The parameters which are not denoted are zero.
 The function $T_\mathrm{CC}$ satisfies the following rule:
 $T_\mathrm{CC}(N_1,N_2,N_3)=T_\mathrm{CC}(N_1,N_2,5)$ if $N_3 > 5$.
}
\label{tb:tbTij}
\begin{ruledtabular}
\begin{tabular}{@{\hspace{2.0em}}ccccl@{\hspace{2.0em}}}
		~ & \multicolumn{3}{c}{Variables}  & ~ \\ 
		\cline{2-4}
		Function & $N_1$ & $N_2$ & $N_3$ & \hspace{2.0em}Value \\ 
		\hline
	$T_\mathrm{CC}(N_1, N_2, N_3)$ & 2& 2& 1& $-$0.070280085 \\ 
	~ & 2& 2& 5& $-$0.00809675 \\ 
\end{tabular}
\end{ruledtabular}
\end{table}

\end{document}